\documentclass[preprint,showpacs,preprintnumbers,amsmath,amssymb,floatfix,aps]{revtex4}

\usepackage{graphicx}
\usepackage{dcolumn}
\usepackage{bm}
\newcommand{\vecvar}[1]{\mbox{\boldmath$#1$}}

\begin{document}

\preprint{PRESAT-8001}

\title{First-principles study of tunnel current between scanning tunneling microscopy tip and hydrogen-adsorbed Si(001) surface}

\author{Tomoya Ono, Shinya Horie, Katsuyoshi Endo, and Kikuji Hirose}
\affiliation{Graduate School of Engineering, Osaka University, Suita, Osaka 565-0871, Japan}

\date{\today}

\begin{abstract}
A scanning tunneling microscopy (STM) image of a hydrogen-adsorbed Si(001) surface is studied using first-principles electron-conduction calculation. The resultant STM image and scanning tunneling spectroscopy spectra are in agreement with experimental results. The contributions of the $\pi$ states of bare dimers to the tunnel current are markedly large, and the $\sigma$ states of the dimers rarely affect the STM images. The tunnel currents do not pass through the centers of the dimers but go through the edges of the dimers with local loop currents. In addition, when the tip exists above the hydrogen-adsorbed dimer, there are certain contributions from the $\pi$ state of the adjacing bare dimers to the tunnel current. This leads to the STM image in which the hydrogen-adsorbed dimers neighboring bare dimers look higher than those surrounded by hydrogen-adsorbed dimers. These results are consistent with the experimental images observed by STM.
\end{abstract}

\maketitle
There is considerable theoretical and experimental interest in the electronic and geometric structures of semiconductor surfaces, especially those of silicon surfaces. Thus far, scanning tunneling microscopy (STM) \cite{stm} has made it possible to observe the structures at surfaces with atomic resolution and has provided much useful information on surface configurations \cite{ref}. It is well known that the STM image is not directly related to the atomic configuration but depends on the electronic structure at the surface. For example, in the case of H-, Cl-, or NH$_3$- adsorbed dimers of Si(001) surfaces, the adsorbed dimers look geometrically lower than bare dimers in STM images \cite{adsorb0,adsorb2, adsorb3,adsorb5}.

On the theoretical side, first-principles calculations based on density functional theory \cite{hkks} have been employed for identification of surface atomic structures in STM images. Analyses on the basis of the local density of states (LDOS) of surfaces can yield reasonable results for STM images under the Tersoff-Hamann (TH) \cite{th} procedure, which means that the variation of the tunnel current with respect to a bias voltage is proportional to the LDOS of the surface at the tip position. By virtue of its simplicity, the TH approximation has been successfully applied to the interpretation of STM images and has provided us with reasonable results in many cases. However, there are some inherent uncertainties in the TH approximation. For example, the images generated by the TH approximation sometimes involve spatially localized surface states \cite{comment0} which do not contribute to electron conduction due to the disconnection with the propagating Bloch states inside electrodes \cite{noguera0-makoshi-kkobayashi}, and there are no definite ways to distinguish actually dominant states to electron conduction from other localized states. In addition, it is difficult to examine the tunnel current path because the global wave functions for infinitely extended states continuing from the substrate to the tip are not accounted for in the framework of the TH approximation. For the accurate interpretation of STM images, it is mandatory to compute the tunnel current flowing between the tip and surfaces.
\begin{figure}[tbh]
\caption{Schematic representation of computational model. The thick solid line represents the supercell. Large light, small light, and dark balls are aluminum, hydrogen, and silicon atoms, respectively. In the top view, atoms are denoted by large and small balls according to the distance from the surface.}
\label{fig:1}
\end{figure}

In this Rapid Communication, the first-principles electron-conduction calculation for tunnel current flowing between the hydrogen-adsorbed Si(001) surface and tip of STM is implemented. To the best of our knowledge, this is the first theoretical indication of the current image and current pass in STM systems using first-principles calculation. The calculated STM image and scanning tunneling spectroscopy (STS) spectra are consistent with experiments. We found that the surface $\pi$ states of the bare dimer largely affect the current image while the contributions of the $\sigma$ states of the dimers are quite small. In addition, tunnel currents flow through the edges of the dimers rather than the centers of the dimers. Moreover, there are certain contributions of the $\pi$ states of the neighboring bare dimer to the tunnel current when the tip is above the hydrogen-adsorbed dimer.
\begin{figure}[tbh]
\caption{Theoretical constant-height STM image of Si(001) surface. In order to compare with STM images, positions with high tunnel current are shaded lighter than those with low tunnel current. The meanings of the symbols are the same as those in Fig.~\ref{fig:1}.}
\label{fig:2}
\end{figure}

Our first-principles calculation program is based on the real-space finite-difference approach \cite{rsfd,rsfdono,icp}, in which boundary conditions are not constrained to be periodic and basis sets used to expand wave functions are not required. Therefore, one can calculate not only electronic structures of surfaces but also tunnel current flowing between a tip and a surface of STM accurately by virtue of making efficient use of the advantages in boundary conditions. Moreover, the wave functions in vacuum regions can be strictly determined in the real-space formalism, while their characteristics unphysically depend on the choice of basis set in the linear combination of atomic orbitals approach. Figure~\ref{fig:1} shows the computational model. The scattering region consists of five silicon layers with the (2$\times$2) lateral cell and pyramidal tip made of aluminum atoms. We adopt the nine-point finite-difference formula for the derivative arising from the kinetic-energy operator of the Kohn-Sham equation in the density functional theory \cite{hkks}. We take a cutoff energy of 23 Ry, which corresponds to a grid spacing of 0.65 a.u., and a higher cutoff energy of 210 Ry in the vicinity of nuclei with the augmentation of double-grid points \cite{rsfdono,icp}. The ion cores are represented by the norm-conserving pseudopotential \cite{ncps95} given by Troullier and Martins \cite{tmpp} and the exchange-correlation effects are treated by the local density approximation \cite{lda}. The supercell is imposed the periodic boundary condition in the [110] and [1$\bar{1}$0] directions while the nonperiodic boundary condition in the [001] direction. Hydrogen atoms are adsorbed on only one silicon dimer and the bottom side of the film remains bare. All of them are suspended between semi-infinite aluminum jellium electrodes. The distance between the tip and surface is set to be $\sim$ 9.5 a.u. so as to save computational cost. The atomic geometry of the silicon surface is determined by first-principles structural optimization using the conventional periodic supercell technique in which the bottom side of the film is terminated by hydrogen atoms and the tip is absence. The global wave functions for infinitely extended states continuing from one electrode side to the other are determined by employing the overbridging boundary matching method \cite{icp,obm}. The tunnel current of the system is evaluated using the Landauer-B\"uttiker formula \cite{landauer}
\begin{equation}
\label{eqn:current}
I=-\frac{2e}{h}\int^{E_F}_E T(E') dE',
\end{equation}
where $h$ is Planck's constant, $e$ is the electron charge, $E_F$ is the Fermi level, and $T(E)$ is the transmission coefficient of incident electrons at energy $E$. The tunnel current is examined on 4$\times$4 points on the surface by displacing the aluminum tip, and the STM image is generated by integration of the tunnel current over the range from the Fermi level $E_F$ to $E_F-1.5$ eV.

Figure~\ref{fig:2} depicts a constant-height STM image generated by the tunnel current. The tunnel current at the bare dimer is larger than that at the hydrogen-adsorbed dimer. Since the tip must approach the sample surface in order to achieve the constant tunnel current, the hydrogen-adsorbed dimer looks geometrically lower than the bare dimer in STM, which agrees with the experimental results \cite{adsorb2,adsorb3,adsorb5}. It is also noted that a tunnel current of $\sim$ $10^{-6}$ A is consistent with the experimental result in consideration of the fact that the tunnel current increases by about seven times per 1 \AA \hspace{2mm} as the tip-surface distance decreases \cite{adsorb2,adsorb5}. One can recognize that the resultant image of hydrogen adsorbed dimer is not consistent with typical experimental image in which the hydrogen adsorbed dimer shows two protrusions across the dimer directions with a minimum at the center \cite{adsorb2}. This is due to the effect of the adjacent bare dimer discussed later \cite{comment1}. Indeed, the two protrusions look fuzzy when the hydrogen adsorbed dimers neighbor to bare dimers in the experimental image.

\begin{figure}[tbh]
\begin{center}
\end{center}
\caption{STS spectra on bare dimer (solid curve) and hydrogen-adsorbed dimer (dashed curve) in the computational model of Fig.~\ref{fig:1}. The zero of energy is chosen to be the Fermi level.}
\label{fig:3}
\end{figure}

In order to explore the relationship between the surface electronic structure and the tunnel current, Fig.~\ref{fig:3} shows the calculated STS spectra when the tip is located above the bare dimer and above the hydrogen-adsorbed dimer using Eq. (\ref{eqn:current}). Considering the fact that the local density approximation underestimates the band gap by $\sim$ 0.6 eV, the peaks in the spectra are in agreement with those in experiments \cite{adsorb3}. The $\pi$ ($\pi^*$) state of the bare dimers is located at about $1.5-0.5$ ($0.5-1.0$) eV below (above) the Fermi level \cite{adsorb3}. Therefore, we can conclude that the peaks in the STS spectra of the bare dimer at $-$0.6 and $0.6$ eV are due to the $\pi$ and $\pi^*$ states, respectively. It is noted that in the case of hydrogen-adsorbed dimer the peaks of $\pi$ and $\pi^*$ are also observed although the dangling bonds of the dimer are passivated.

For further interpretation of the STM image and STS spectra, we show in Fig.~\ref{fig:4} the charge distributions and the current distributions of electrons incident from the silicon surfaces to the tip with energies between the Fermi level $E_F$ and $E_F-1.5$ eV, and the contour plot of the LDOS which are obtained using the conventional three-dimensionally periodic supercell for the same energy range. The current distribution is derived by the equation of continuity. From Fig.~\ref{fig:4}(a) in which electrons accumulate above and below the Si-Si bond of the bare dimer, one can recognize that the $\pi$ bonding state largely contributes to the tunnel current although there is little contribution of the $\sigma$ bonding state; the surface $\pi$-states of the bare dimer, which are expected to be localized in the surface region, actually contribute to electron conduction from the sample electrode to the tip electrode in the case of the Si(001) surface.

In addition, the local loop currents rotating around the atoms consisting the dimers are observed in Fig.~\ref{fig:4}(b). These peculiar characteristics of the tunnel current can be explained by the decomposition of the scattering wave functions into the atomic orbitals around the respective atoms,
\begin{equation}
\psi_E(\vecvar{r})=C^s_E\phi_s(\vecvar{r})+ \sum_{\alpha=x,y,z}C^{p_\alpha}_E\phi_{p_\alpha}(\vecvar{r}).
\label{eqn:expand}
\end{equation}
Here, $x$, $y$ and $z$ directions correspond to [110], [1$\bar{1}$0] and [001] directions, respectively. By substituting Eq. (\ref{eqn:expand}) into the equation of continuity, we have
\begin{eqnarray}
\lefteqn{\vecvar{I}(\vecvar{r})}\nonumber \\
&=& -\frac{eh}{\pi} \Biggl[ \sum_{\alpha=x,y,z} \int_E^{E_F} \mbox{Im}[C^{s,*}_{E'})C^{p_\alpha}_{E'} \phi_s^*(\vecvar{r}) \vecvar{\nabla} \phi_{p_\alpha}(\vecvar{r})] dE' \nonumber \\
&& + \sum_{\alpha=x,y,z} \int_E^{E_F} \mbox{Im}[C^{p_\alpha,*}_{E'}C^s_{E'} \phi_{p_\alpha}^*(\vecvar{r}) \vecvar{\nabla} \phi_s(\vecvar{r})] dE' \nonumber \\
&& + \sum_{\stackrel{\alpha,\beta=x,y,z}{\alpha\ne\beta}} \int_E^{E_F} \mbox{Im}[(C^{p_\beta,*}_{E'})C^{p_\alpha}_{E'} \phi_{p_\beta}^*(\vecvar{r}) \vecvar{\nabla} \phi_{p_\alpha}(\vecvar{r})] dE'\Biggr] .
\label{eqn:expand1}
\end{eqnarray}
The first term of Eq. (\ref{eqn:expand1}) corresponds to currents flowing along the $\alpha$ direction and the second one to local loop currents rotating on the $\alpha$-$\beta$ plane. Due to the geometrical symmetry of atoms, the currents except that flowing along the $z$ direction vanish inside the substrate. On the other hand, the current rotating on the $x$-$z$ plane emerges on the surface plane because the atoms on the surface plane form dimers. Consequently, the currents at the edges of the dimers are enhanced as seen in Fig.~\ref{fig:4}(b). The wave functions of the surface can be also expressed in terms of the following basis set,
\begin{eqnarray}
\phi_{\pi}&=&[\phi_s(\vecvar{r})+\phi_{p_z}(\vecvar{r})]/\sqrt{2} \nonumber \\
\phi_{\sigma_D}&=&\phi_{p_x}(\vecvar{r}) \nonumber \\
\phi_{\sigma_{B1}}&=&[-\phi_s(\vecvar{r})+\sqrt{2}\phi_{p_y}(\vecvar{r})+\phi_{p_z}(\vecvar{r})]/2 \nonumber \\
\phi_{\sigma_{B2}}&=&[-\phi_s(\vecvar{r})-\sqrt{2}\phi_{p_y}(\vecvar{r})+\phi_{p_z}(\vecvar{r})]/2.
\end{eqnarray}
One can notice that the $\pi$ state contributes to the current along the $z$ direction while the $\sigma$ state affects the local loop current around the surface atoms. This agrees with the STS spectra of Fig.~\ref{fig:3} and the charge distributions of Fig.~\ref{fig:4}(a), in which the $\pi$ states actually contribute to the tunnel current.

\begin{figure*}[tbh]
\caption{(Color) (a) Charge distributions and (b) current distributions, $-\vecvar{I}(\vecvar{r})/e$, of incident electrons from the silicon surfaces to the tip. The aluminum tip is located above the bare dimer in the upper panel and above the hydrogen-adsorbed dimer in the lower pannel. The density distributions of electrons incident from the bottom electrode with energies between E$_F$ and $E_F-1.5$ eV are depicted. (c) Contour plots of the LDOS, which are computed using the conventional three-dimensionally periodic supercell, for the same energy ranges as those in (a). The channel electron distributions and the contour plots of the LDOS are illustrated on a logarithmic scale and each contour represents twice or half the density of the adjacent contour lines. The map is along the cross section indicated by the dotted lines of A, B, or C in Fig.~\ref{fig:1}. Gray, pink, and blue balls are aluminum, hydrogen, and silicon atoms, respectively. The large (small) balls indicate the atomic positions on (above) the cross section.}
\label{fig:4}
\end{figure*}

Moreover, it is observed in the lower panel of Fig.~\ref{fig:4}(b) that there is a certain contribution from the $\pi$ state of the neighboring bare dimer to the tunnel current, which agrees with the STS spectra of Fig.~\ref{fig:3} where the $\pi$ state affects the spectra even when the tip locates above the hydrogen-adsorbed dimer. In order to estimate the contribution of the neighboring bare dimer, we evaluate the following quantity, $R=\int_{S_{bar}} I_z(\vecvar{r}) dS/\int_{S_{ads}} I_z(\vecvar{r}) dS$, where the surface integral regarding $S_{bar}$ ($S_{ads}$) is implemented on the rectangle (2$\times$1) plane just above the bare (hydrogen-adsorbed) dimer. The $R$ is $\sim$ 0.25 around the bisection of the surface-dimer- and tip-apex-atom planes. When the tip-surface distance increases to 10 \AA, the $R$ becomes 0.53 at the bisection plane, which implies that 35 \% of the tunnel current is attributed to the neighboring bare dimer: The transmission from the $\pi$ states to the tip reduces more moderately than that from the states around the hydrogen-adsorbed dimer as the tip-surface distance increases because the $\pi$ state is widely expended to the vacuum region. In addition, our result is also consistent with the height profile of Ref. \cite{adsorb2} in which hydrogen-adsorbed dimers neighboring bare dimes look geometrically higher by $\sim$ 0.25 \AA \hspace{2mm} than those surrounded by hydrogen-adsorbed dimers in the [010] direction.

Let us discuss the relationship between the charge distribution of the incident electrons and the LDOS. In the case of the bare dimer, there is no significant difference in characteristics of the counter plot of LDOS in Fig.~\ref{fig:4}(c) from that of the charge distribution of incident electrons in Fig.~\ref{fig:4}(a). Although a certain influence of the adsorbed hydrogen atoms on the LDOS is observed around the Si-H bonds in Fig.~\ref{fig:4}(c), the contributions of the states affected by the hydrogen atoms to the distribution of incident electrons are small compared with the Si-Si bonding state in Fig.~\ref{fig:4}(a); the densities around the hydrogen atoms and Si-H bonds are one forth of that around the Si-Si bond in Fig.~\ref{fig:4}(a) while that is half in Fig.~\ref{fig:4}(c). We assured that this situation does not change in the case of large tip-substrate distance. This result implies that by being adsorbed hydrogen atoms the orbitals comprised by the $s$ and $p_z$ orbitals of the surface atoms couple weakly with the waves continuing from the substrate compared with those composed by the $p_x$ orbital.

In summary, we have studied an STM image of hydrogen-adsorbed Si(001) surfaces using first-principles electron-conduction calculation. The tunnel current flowing between a hydrogen-adsorbed dimer and the STM tip is smaller than that flowing between a bare dimer and the tip. The $\sigma$ states of the dimer hardly affect the STM image while the contributions of the $\pi$ bonding states to the tunnel current between the bare dimer and the tip are large. In addition, the currents passing through the edges of the dimers are due to the contributions of the $\sigma$ states of the dimers. Since the tunnel current with the tip being above the hydrogen-adsorbed dimer is attributed to the $\pi$ states of the adjacent bare dimers, hydrogen-adsorbed dimers neighboring bare dimers appear geometrically higher than those surrounded by hydrogen-adsorbed dimers in the [010] direction. These results are consistent with experiments.

This research was partially supported by a Grant-in-Aid for the 21st Century COE ``Center for Atomistic Fabrication Technology'', by a Grant-in-Aid for Scientific Research in Priority Areas ``Development of New Quantum Simulators and Quantum Design'' (Grant No. 17064012) and also by a Grant-in-Aid for Young Scientists (B) (Grant No. 17710074) from the Ministry of Education, Culture, Sports, Science and Technology. The numerical calculation was carried out by the computer facilities at the Institute for Solid State Physics at the University of Tokyo and the Information Synergy Center at Tohoku University.


\begin{thebibliography}{}
\bibitem{comment1}
The usage of small supercell results in the discrepancy between the theoretical and experimental. When a plain TH approximation is employed using the supercell with the same lateral length, two clear protrusions can not be observed. If we adopt larger supercell or the surfaces which are fully hydrogenated, we can obtain the images which are consistent with experimental ones. Since our interest here is the tunnel current from the surface in which both bare and hydrogenated dimers exist, we employ the model where one dimer is bare and the other is hydrogenated.
\end{thebibliography}
\end{document}